\documentclass{ptapap}

\author{R. Smolec}[smolec@camk.edu.pl,CAMK]
\author{P. Moskalik}[CAMK]
\author{N.R. Evans}[SAO]
\author{A.F.J. Moffat}[AFM]
\author{G.A. Wade}[GW]
\author{BEST}[BEST]
\affil[CAMK]{Nicolaus Copernicus Astronomical Center, Bartycka 18, 00--716 Warszawa, Poland}
\affil[SAO]{Smithsonian Astrophysical Observatory, MS 4, 60 Garden Street, Cambridge, MA 02138, USA}
\affil[AFM]{D\'epartement de physique, Universit\'e de Montr\'eal, C.P. 6128, Succursale center-Ville, Montr\'eal, Qu\'ebec, H3C 3J7, Canada}
\affil[GW]{Dept. of Physics, Royal Military College of Canada, PO Box 17000, station Forces, Kingston, Ontario, Canada K7K 4B4}
\affil[BEST]{BRITE Executive Science Team}

\title{BRITE observations of classical Cepheids}

\begin{document}

\maketitle

\begin{abstract}
We briefly summarize the BRITE observations of classical Cepheids. Possible detection of modulation in a fundamental mode Cepheid, T~Vul, and of additional non-radial modes in first overtone Cepheids, DT~Cyg and V1334~Cyg, are reported.
\end{abstract}

\section{Introduction}

Only a few classical Cepheids were observed from space so far: Polaris with WIRE and SMEI \citep{bruntt,spreckley}, V1154 Cyg with {\it Kepler} \citep[e.g.][]{szabo1154,derekas1154_2}, RT~Aur and SZ~Tau with {\it MOST} \citep{evans} and 7 Cepheids with {\it CoRoT} \citep{poretti} (see also Plachy, these proceedings). Primary targets for BRITE constellation \citep[see][]{weiss,pablo} are brighter than $4$ mag in $V$, however fainter stars with slow variability can be observed at high precision as well, which we demonstrate in this contribution. For 24 classical Cepheids, the apparent magnitude in the $V$ band is brighter than $6.5$ mag at minimum brightness and these stars were proposed as primary Cepheid targets for BRITE. So far, seven of these were observed, and here we report the initial results of data analysis.

\section{Data analysis}

Basic information on the seven observed Cepheids and the available photometric data  are collected in Tab.~\ref{tab}. Cepheids were observed mostly with the red filter by UniBRITE (UBr), BRITE-Toronto (BTr) and BRITE-Heweliusz (BHr) and with the blue filter by BRITE-Lem (BLb). Cepheids are much fainter in blue; only short photometric series were gathered with this filter (for $\delta$~Cep, DT~Cyg and V1334~Cyg), covering no more than 3 pulsation cycles and the analysis confirms very low precision of the observations. Consequently, we focus on observations obtained with the red filter only. In addition, it turned out that the photometric precision of data for fainter Cepheids collected by UBr is much lower (not better than ground-based photometry for the analysed stars) than that of the data collected with two other `red' satellites. Consequently, we will not discuss the UBr data for T~Vul, DT~Cyg and V1334~Cyg, for which much better BTr photometry is available.

Before the analysis, raw data were processed following the guidelines summarized in the ``BRITE cookbook'' (by A. Pigulski, \textsf{http://brite.craq-astro.ca/doku.php?id=cookbook}). In a nutshell, after initial cleaning of the data (removing of severe outliers and bad orbits), the dominant variability (modelled with Fourier series, as is common for classical pulsators) was removed from the data. Then, the residuals were used to decorrelate the data with CCD temperature, position of the star within the raster, and with orbital phase. Finally, owing to relatively slow variability of Cepheids, we averaged the data gathered over $\sim$20 minutes during each BRITE orbit. Iterative $3\sigma$ clipping was also applied at this stage to remove the outliers within each orbit. Data prepared this way were subject to analysis with the standard consecutive prewhitening technique. We note that original data were delivered in a few separate files (`setups') which we analysed separately. 

\begin{figure}
\includegraphics[width=\textwidth]{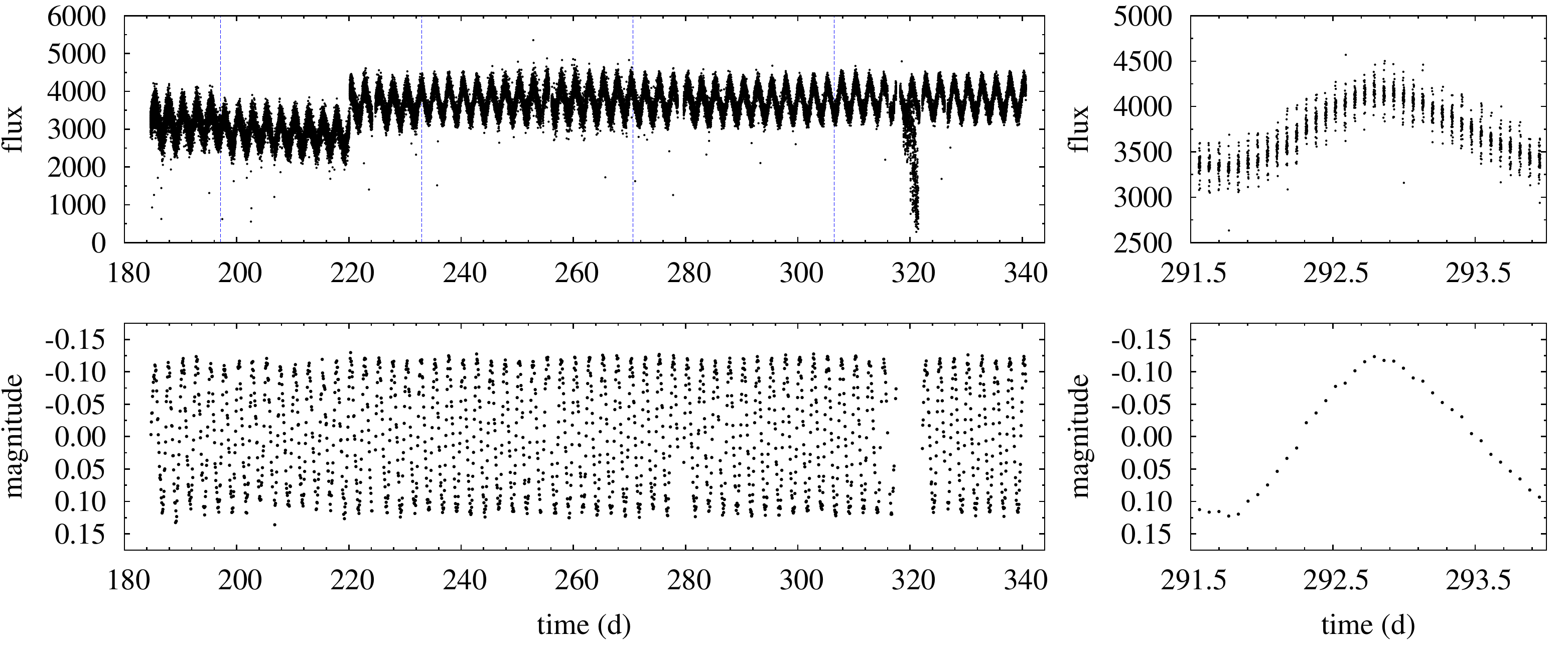}
\caption{Original flux data (top panels; five separate setups as indicated by vertical lines) and final, reduced, magnitude data (bottom panels) for DT~Cyg observed with BTr during the Cyg-II campaign. Right panels show zooms into short section of the data.}
\label{fig:da}
\end{figure}

To illustrate the described procedure, in Fig.~\ref{fig:da} we show the raw flux data for DT~Cyg (top panel; five setups marked with vertical dashed lines) and the final magnitude data (bottom panel). Severe outliers are clearly visible in the raw data and were removed during the analysis. The long flux drop at $t\approx 320$\thinspace d was entirely cut out from the data. A flux jump at $t\approx 220$\thinspace d (within setup two) was caused by significant slowing of the CCD read-out time at that moment (which was applied to reduce the Charge Transfer Inefficiency (CTI) -- see \cite{pablo} for more details). This change in the observing procedure resulted in significant improvement of the data quality -- the photometric dispersion is nearly a factor two smaller after the CTI was resolved. The effect is most pronounced for DT~Cyg and V1334~Cyg (observed during the Cyg-II campaign by BTr); in the analysis of these stars we simply drop the first part of the data (before  $t\approx 220$\thinspace d). As a result, the noise level in the Fourier transform drops significantly (at the cost of decreasing the frequency resolution). The right panels of Fig.~\ref{fig:da} show zooms into small sections of the data. Data gathered over individual orbits are clearly visible in the top panel. These are averaged to a single measurement (bottom panel).

\begin{table}
\label{tab}
\begin{tabular}{lrrrrl}
star & HD & mode & $P$ (d) & $\langle V\rangle$& summary of observational data\\
\hline
X~Sgr        & 161592 &  F & 7.0128 & 4.55 & UBr (Sgr-I)\\ 
W~Sgr        & 164975 &  F & 7.5949 & 4.67 & UBr (Sgr-I)\\ 
T~Vul        & 198726 &  F & 4.4355 & 5.75 & BTr (Cyg-I, Cyg-II)\\
$\delta$~Cep & 213306 &  F & 5.3663 & 3.95 & BHr, BTr (Cas-Cep-I)\\ 
DT~Cyg       & 201078 & 1O & 2.4991 & 5.77 & BTr (Cyg-I, Cyg-II)\\
V1334~Cyg    & 203156 & 1O & 3.3328 & 5.87 & BTr (Cyg-I, Cyg-II)\\
MY~Pup       &  61715 & 1O & 5.6953 & 5.68 & BHr ($\beta$~Pic-I)\\ 
\hline
\end{tabular}
\caption{Basic data about Cepheids observed with BRITE (pulsation mode, pulsation period, mean $V$-band brightness) and indication of satellites and observing campaigns in which data were gathered (red filter only; see \textsf{http://brite.craq-astro.ca/} for more details).}
\end{table}

\section{Results: fundamental mode Cepheids}
\begin{figure}[t]
\includegraphics[width=\textwidth]{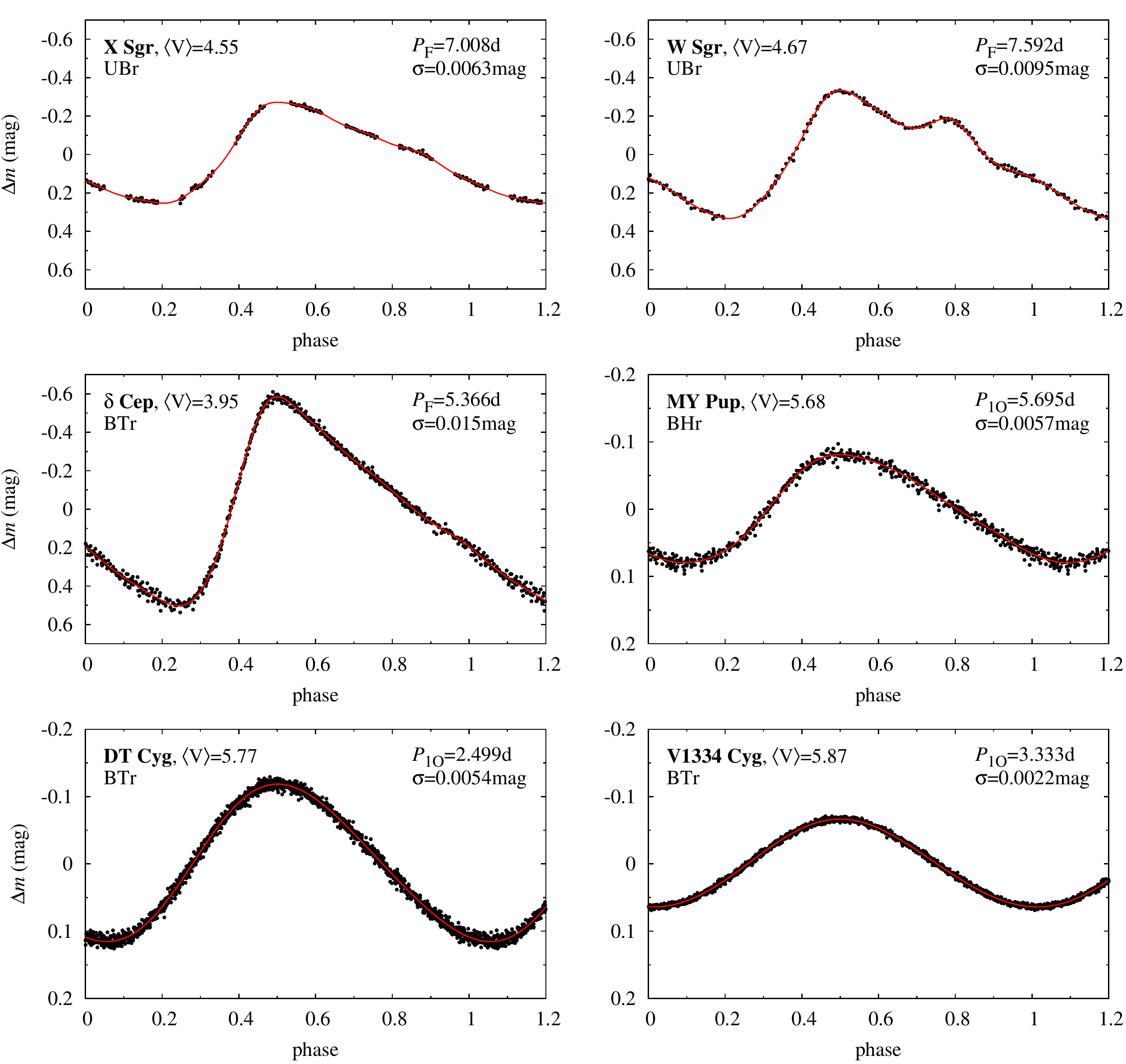}
\caption{Phased light curves with Fourier fits for all but one Cepheids observed with BRITE-Constellation. For T~Vul light curve -- see Fig.~\ref{fig:tvul}. Basic data about Cepheids is given in each panel. Note the different vertical axis range for fundamental and first overtone stars.}
\label{fig:lc}
\end{figure}

\noindent{\bf X Sgr and W Sgr.} For these two fundamental mode pulsators only data from UBr is available and it is only 30\thinspace d long, which corresponds to $\sim$4 pulsation cycles, with incomplete phase coverage. With these data we are able to show the phased light curves only -- Fig.~\ref{fig:lc}. Both stars are bump Cepheids.
\smallskip

\begin{figure}
\includegraphics[width=\textwidth]{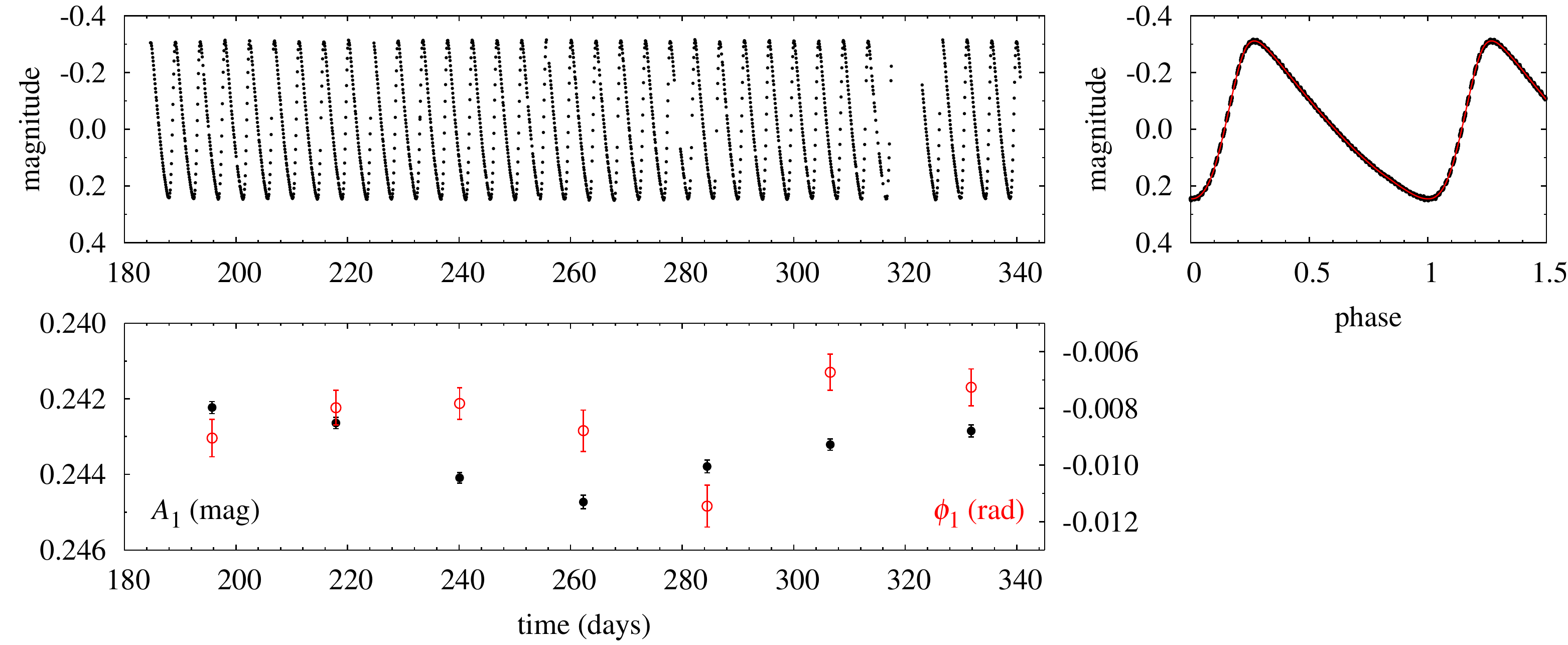}
\caption{Data for T~Vul gathered by BTr during the Cyg-II campaign (top left). Top right panel shows phased light curve with Fourier fit overplotted. Bottom left panel shows the time variation of Fourier amplitude, $A_1$, and Fourier phase, $\phi_1$, obtained with the time-dependent Fourier analysis.}
\label{fig:tvul}
\end{figure}

\noindent{\bf T Vul.} Long and excellent quality BTr observations are available. Although data gathered during the Cyg-I campaign cover only 80 d with several gaps, data gathered during the Cyg-II campaign are 150 d long with nearly 100 per cent duty cycle -- top left panel of Fig.~\ref{fig:tvul}. The phased light curve is plotted in the top right panel of the same figure. The photometric dispersion is very low, $\sigma=0.0021$ mag. In the frequency spectrum of the residual data we detect unresolved power at the frequency of the fundamental mode, which indicates that the amplitude and/or phase may vary in time. Time-dependent Fourier analysis \citep[see][]{kbd87} indeed shows -- see bottom panel of Fig.~\ref{fig:tvul} -- that pulsation of T~Vul may be modulated: the pulsation amplitude and pulsation phase vary smoothly in time. More observations are needed however to resolve the suspected modulation. 
\smallskip

\noindent{\bf $\boldmath{\delta}$~Cep.} Data from BTr and BHr are available; the latter are of inferior quality however, and we will not discuss them here. $\delta$~Cep is the brightest star in our sample and was observed with BTr, which collected top-quality photometry for the much fainter star T~Vul, as we have just discussed. For an as-yet unidentified reason, the $\delta$~Cep photometry obtained by BTr is poor. It also only covers 9 pulsation cycles. For this star, we show only the phased light curve (Fig.~\ref{fig:lc}) and note that revisiting the star is necessary. In the frequency spectrum, an increased power is noted at the expected location of the radial first overtone, but the detection is rather weak. In addition, the frequency resolution is low, and the frequency spectrum, particularly at the low frequency range, is somewhat sensitive to various reduction schemes we tried. Much better and longer observations are needed to confirm the detection.

\section{Results: first overtone Cepheids}

\noindent{\bf MY~Pup.} Data from BHr are available and cover 14 consecutive pulsation cycles. However, for the first half, the data are of significantly lower quality. The phased light curve is presented in Fig.~\ref{fig:lc}. No significant signal is detected in the frequency spectrum, except for the first overtone and its harmonics.
\smallskip

\begin{figure}
\includegraphics[width=\textwidth]{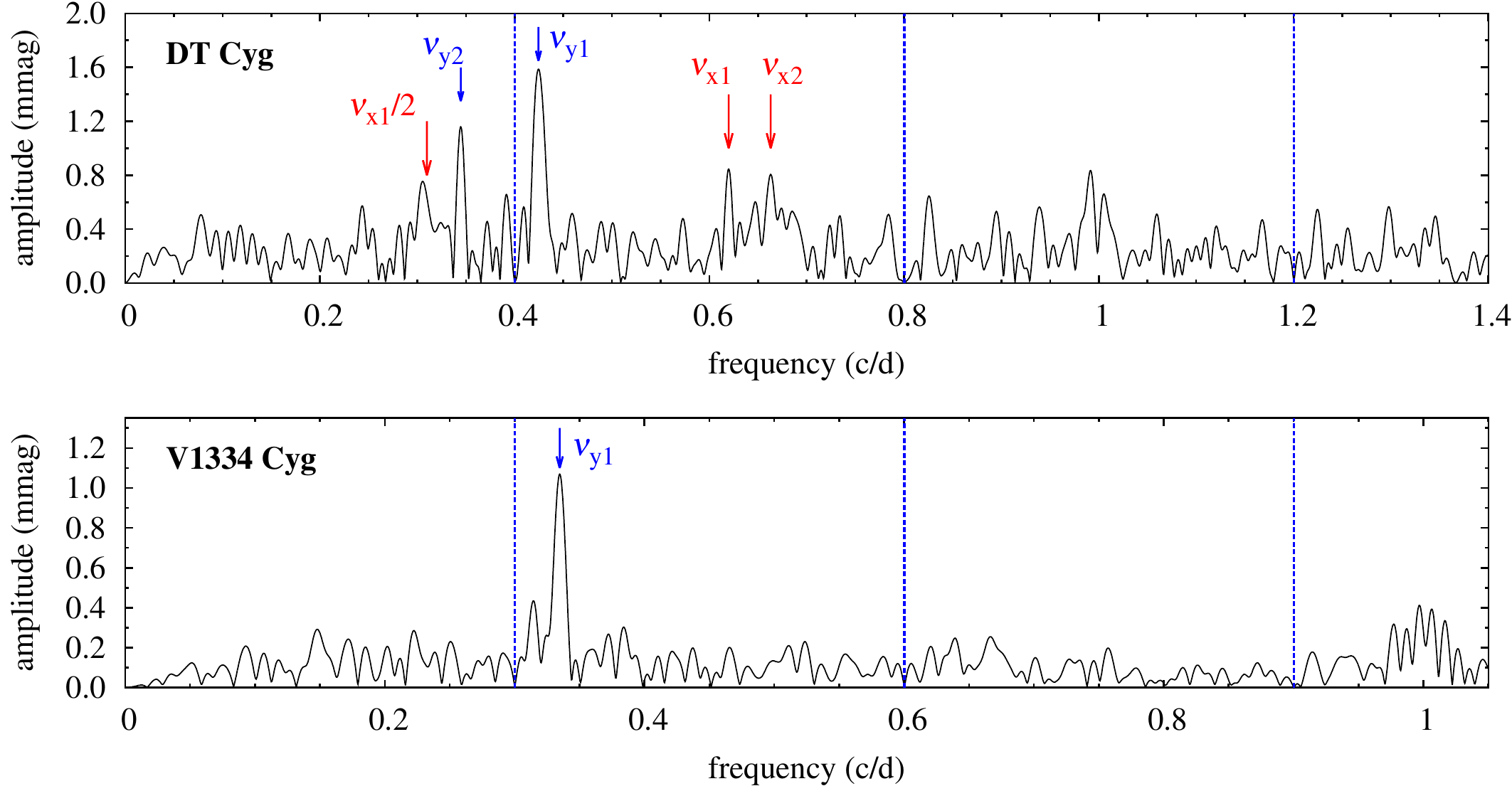}
\caption{Frequency spectra for DT~Cyg (top panel) and V1334~Cyg (bottom panel) after prewhitening with the first overtone and its harmonics (dashed lines).}
\label{fig:1Ofsp}
\end{figure}

\noindent{\bf DT~Cyg.} Just as for T~Vul, data gathered during the Cyg-I campaign are shorter and of inferior quality (due to data gaps). Here we focus on data gathered during the Cyg-II campaign. The phased light curve is presented in Fig.~\ref{fig:lc} and the frequency spectrum of the data, after prewhitening with the first overtone frequency, $\nu_1$, and its harmonics, is plotted in the top panel of Fig.~\ref{fig:1Ofsp}. Close to $\nu_1$, two significant peaks ($\nu_{\rm y1}$ and $\nu_{\rm y2}$) are detected. Period ratios (with the radial first overtone) are $P_{\rm y1}/P_1=0.943$ and $P_{\rm y2}/P_1=1.161$, and thus additional periodicities cannot correspond to radial modes. They may be due to non-radial pulsation or due to modulation -- see \cite{mk09} for similar cases among first overtone Cepheids in the Magellanic Clouds. In Fig.~\ref{fig:1Ofsp} we also marked three other peaks. Although they are not pronounced, their location is telling: for period ratios we have $P_{\rm x1}/P_1=0.646$ and $P_{\rm x2}/P_1=0.604$. Signals with similar period ratios are quite common in first overtone Cepheids -- see e.g. \cite{smc_cep,ss16}. In the Petersen diagram they form three sequences -- the two periodicities in DT~Cyg would correspond to the top and the bottom sequence. According to the model proposed by \cite{wd16}, these are harmonics of the non-radial $\ell=7$ and $\ell=9$ modes. A weak signal at $\nu_{\rm x1}/2$ (also marked in Fig.~\ref{fig:1Ofsp}) would then correspond to the non-radial, $\ell=7$ mode. We note that in the frequency spectrum of data gathered during the Cyg-I campaign the signals at $\nu_{\rm y1}$ and $\nu_{\rm x1}/2$ were also detected. Other signals were not detected, but the overall noise level is significantly larger for the Cyg-I data.
\smallskip

\noindent{\bf V1334~Cyg.} Similar to the case of DT~Cyg and T~Vul, we focus on the data gathered during the Cyg-II campaign. The phased light curve is presented in Fig.~\ref{fig:lc}, while the frequency spectrum, after prewhitening with the first overtone and its harmonics, is plotted in the bottom panel of Fig.~\ref{fig:1Ofsp}. Close to $\nu_1$ pronounced peak is detected. The period ratio (with the radial first overtone) is $P_{\rm y1}/P_1=0.895$, and is inconsistent with simultaneous excitation of two radial modes. The additional signal may be due to a non-radial mode or due to modulation. Significant signal with the same frequency was also detected in the data gathered during the Cyg-I campaign.

\acknowledgements{Based on data collected by the BRITE Constellation satellite mission, designed, built, launched, operated and supported by the Austrian Research Promotion Agency (FFG), the University of Vienna, the Technical University of Graz, the Canadian Space Agency (CSA), the University of Toronto Institute for Aerospace Studies (UTIAS), the Foundation for Polish Science \& Technology (FNiTP MNiSW), and National Science Centre (NCN). This research is supported by the Polish National Science Centre through grants DEC-2012/05/B/ST9/03932 and 2011/01/M/ST9/05914. GAW is supported by an NSERC Discovery Grant.}

\bibliographystyle{ptapap}
\bibliography{britecep}

\end{document}